\documentclass{elsart}

 \usepackage{epsfig}

\usepackage{amssymb}

\begin{document}

\begin{frontmatter}



\title{Pseudospin, supersymmetry and the shell structure of atomic nuclei}


\author{S. Typel}

\address{Grand Acc\'{e}l\'{e}rateur National d'Ions Lourds (GANIL),
 CEA/DSM-CNRS/IN2P3, BP 55027, F-14076 Caen Cedex 5, France}

\begin{abstract}
The evolution of single-particle energies with varying isospin asymmetry
in the shell model is an important issue when 
predicting changes in the shell structure for exotic nuclei. 
In many cases pseudospin partner levels, that are almost
degenerate in energy for stable nuclei, 
are relevant in extracting the size of the shell gaps.
A breaking of the pseudospin symmetry can affect the size 
of these gaps and change the magic numbers accordingly.
The strength of the pseudospin splitting is expected to depend in particular
on isovector-dependent and tensor contributions
to the effective nuclear interaction.
A description employing supersymmetric quantum mechanics allows
to derive a pseudospin symmetry breaking potential that is
regular in contrast to the pseudospin-orbit potential
in the conventional relativistic treatment. 
The derived perturbation potential provides a 
measure to quantify the symmetry breaking and it can be employed
to improve mean-field calculations in order
to better reproduce the experimentally observed shell evolution.
General potentials with exact pseudospin symmetry are obtained
that can be used in relativistic mean-field Hamiltonians.
\end{abstract}

\end{frontmatter}




\section{Introduction}

\begin{figure}[h]
\center{\includegraphics[width=12cm,angle=0]{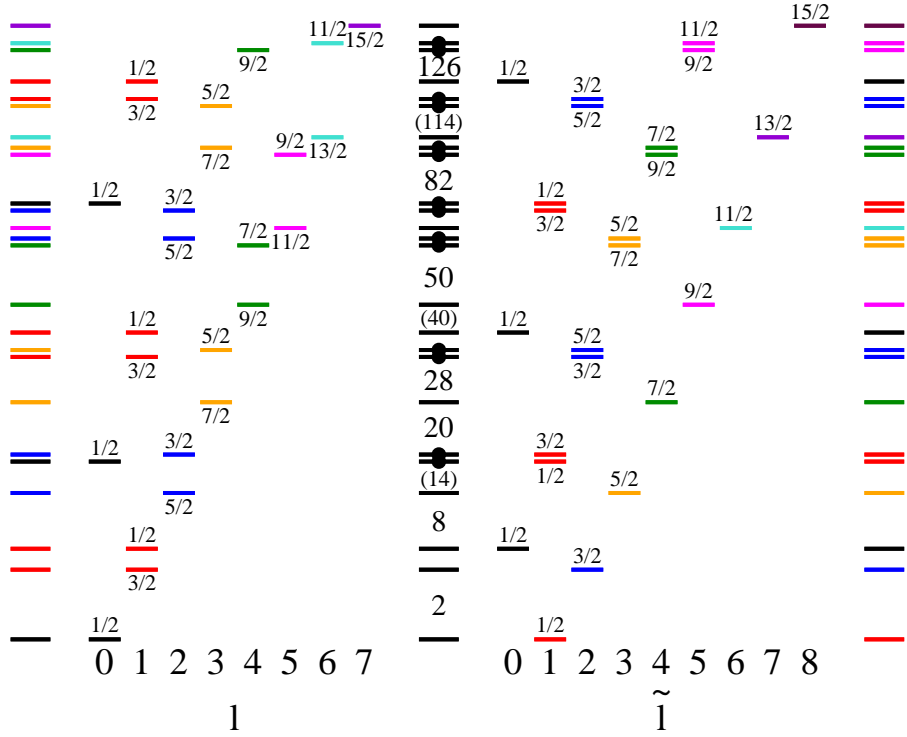}}
\caption{\label{fig:01}
Single-particles states in the conventional shell model description
with spin-orbit partners $j=l\pm 1/2$ for given orbital angular momentum
$l$ (left) and in the supersymmetric approach with
pseudospin partners $j=\tilde{l}\pm 1/2$ for given pseudo-angular momentum
$\tilde{l}$ (right). The pseudospin partner states are marked with
a solid circle in the central level scheme. Additionally the magic numbers 
for shell closures are given. Subshell closures are enclosed in
parentheses.}
\end{figure}

The single-particle shell model with strong
central and spin-orbit potentials is a cornerstone of nuclear 
structure physics since about sixty years. It was introduced
by  Haxel, Jensen and Sue\ss{} \cite{Hax48,Sue49,Jen49,Hax49}
and Goeppert-Mayer \cite{Goe49} in order to
explain the experimentally observed occurrence 
of magic numbers for nuclei close to the valley of $\beta$-stability.
In this picture,
nucleons move as independent particles in a spherical mean-field. 
The central potential usually has a shape 
similar to the nuclear density distribution and the spin-orbit potential
is peaked at the nuclear surface.
The large spin-orbit potential explains the observed 
energy splittings between two spin-orbit partners
with total angular momentum $j_{> \atop <}=l \pm 1/2$ 
and parity $\pi = (-1)^{l}$
for a given orbital angular
momentum $l$, see the left part of figure \ref{fig:01}. 
This essential feature of the model
generates the particular shell structure with the appearance of major
shell closures at nucleon numbers 2, 8, 20, 28, 50, 82, 126, \dots
and subshell closures at, e.g., 14, 40, 114. The single-particle
wave functions of the mean-field calculation can
serve as a basis in order to construct fully antisymmetrized
many-body wave functions. Using a suitably chosen subset 
in the many-body space
with an appropriate residual interaction, the excitation
spectrum of a nucleus can be described with high accuracy by diagonalizing
the corresponding many-body Hamiltonian, see, e.g., ref.\ \cite{Cau05}. 

Experiments with exotic nuclei suggest a change of
the well-known shell structure outside the valley of stability
and the appearance of new (sub-)shell closures often attributed
to changes in the spin-orbit splitting or halo effects
\cite{Gui84,Huc86,Kel94,Mot95,Sch96,Gla97,Sim99,Oza00,Gra01,Pri01,%
Soh02,Jan02,Dlo03,Sch04,Din05,For05,Gra05,Gau06,Rej07,Gad08}.
It is very important to understand the evolution of
the single-particle energies with a change in isospin asymmetry and to 
identify the origin of this effect. A change in the size of shell gaps
can have important consequences in r-process
nucleosynthesis \cite{Che95,Pea96,Kra98,Kra07,Mar07}.
In recent years, contributions
to the nucleon-nucleon potential that change the usual shell structure
when the isospin asymmetry of the system increases
have been discussed extensively in the literature 
\cite{Dob94,Lal98,Ots01,Pea01,Rin02,%
Ots02,Ots03,Ots04,Ots05,Ots06,Ots07,Dob07,Col07}.
There are mainly two contributions to the nucleon-nucleon
potential that have been considered:
1.\ a central interaction of Majorana-type $V_{M}(r)
\vec{\sigma}_{1} \cdot \vec{\sigma}_{2}
\vec{\tau}_{1} \cdot \vec{\tau}_{2}$ (spin-isospin interaction)
and 
2.\ a non-central isospin-tensor interaction
$V_{IT}(r)S_{12}\vec{\tau}_{1} \cdot \vec{\tau}_{2}$ with the tensor
operator $S_{12} = 3 (\vec{\sigma}_{1}\cdot\vec{r}_{1})
(\vec{\sigma}_{2}\cdot\vec{r}_{2})/(r_{1}r_{2}) - 
\vec{\sigma}_{1}\cdot \vec{\sigma}_{2}$,
 where the absolute strength of the
potentials depends on the particular radial functions 
$V_{M}(r)$ and $V_{IT}(r)$ with $r=|\vec{r}_{1}-\vec{r}_{2}|$.
The monopole part of these interactions 
is found to be strongly attractive between
protons and neutrons when the spins of the two nucleons are anti-aligned,
i.e., for $\pi l_{j_{<}}$ and $\nu l^{\prime}_{j^{\prime}_{>}}$
(or $\pi l_{j_{>}}$ and $\nu l^{\prime}_{j^{\prime}_{<}}$)
with $j_{<} = l - 1/2$, $j^{\prime}_{>} = l^{\prime}+1/2$ 
(in case of the central
spin-isospin interaction one has $l=l^{\prime}$). Hence, the filling of one
level with protons (neutrons) will lead to a strong shift of the
corresponding neutron (proton) states changing the spin-orbit splittings and,
depending on the involved levels, the size of the shell gaps.


Of particular importance is
the behaviour of the levels that define the conventional shell gaps.
It is interesting to note that for the (sub-)shell closures at
14, 20, 28, 50, 82, 114 and 126 at least one level defining the gap
belongs a pair of so-called pseudospin partners.  These states are 
marked by solid circles in the central level scheme in figure \ref{fig:01}.
The total and orbital angular quantum numbers of the partner states are related
to the so-called pseudo orbital angular momentum $\tilde{l}$ by
$j_{<}=\tilde{l}-1/2$, $l_{<}=\tilde{l}-1$ and 
$j_{>}=\tilde{l}+1/2$, $l_{>}=\tilde{l}+1$,
respectively. Both levels have the same parity $\pi = (-1)^{\tilde{l}+1}$.
In addition, the principal quantum number $n$ of the state
with $j_{<}=\tilde{l}-1/2$ is larger than the principal quantum number 
of the state with $j_{>}=\tilde{l}+1/2$ by one unit. For example, the doublets
$\{2s_{1/2},1d_{3/2}\}_{so}=\{1\tilde{p}_{1/2},1\tilde{p}_{3/2}\}_{ps}$ 
with $\tilde{n}=1$, $\tilde{l}=1$ and
$\{2p_{3/2},1f_{5/2}\}_{so}=\{1\tilde{d}_{3/2},1\tilde{d}_{5/2}\}_{ps}$ 
with $\tilde{n}=1$, $\tilde{l}=2$ are pseudospin partners.
The indices $so$ and $ps$ denote the spin-orbit and pseudospin representation
of the states, respectively.
Many partner levels of pseudospin pairs are experimentally found
to be almost degenerate in energy suggesting the existence of
a particular symmetry
in the mean-field Hamiltonian for the single-particle states. 
This feature has been noticed already more than forty years ago 
and the concept of pseudospin and pseudo-orbital angular momentum
was introduced \cite{Hec69,Ari69}.
A breaking of the pseudospin symmetry in stable and exotic nuclei 
is potentially important for a change of the level structure and the
appearance of new magic numbers.
It is important to identify the contributions to the nuclear interaction
that are responsible for a breaking or the restoration of the pseudospin
symmetry. To this end, a quantitative measure of the pseudospin 
splitting is needed that can be used to compare and improve theoretical models.

Bohr, Hamamoto and Mottelson \cite{Boh82} discussed a transformation 
of the form $\vec{\sigma} \cdot \vec{r} / r$ that connects the 
angular momentum part of the pseudospin partner levels (see eq.\
(\ref{eq:sigmadotr})). However, it cannot account for the change of
the principal quantum number $n$. Hence, a dynamical symmetry involving
the momentum operator $\vec{p}$ is required. Also it was observed
that the SU(3) and pseudo-SU(3) states of the shell model 
are related by transformations that are elements of the orthosymplectic
superalgebra Osp(1/2) \cite{Cas92,Bal92}. 
The many-particle 
pseudo-SU(3) theory is successfully used in the description of
heavy deformed nuclei in the context of the interacting boson model (IBM)
applying supersymmetry, see, e.g., ref.\ \cite{Iac93}. Eventually, the
relevant momentum helicity operator $\vec{\sigma} \cdot \vec{p} / p$
was identified that accomplishes the pseudospin transformation
and it was noticed that
the Dirac equation of free or in-medium nucleons (in the 
Dirac-Brueckner approach) is invariant under this
transformation \cite{Blo95,Blo96}.
The essential step in explaining the pseudospin symmetry
was the application of a relativistic description.
In this approach, e.g., in the form of
relativistic mean-field (RMF) models,
the single-particle states are described by Dirac spinor wave functions
with large upper and small lower components. Ginocchio observed that
the second-order differential equation for the lower component
is approximately invariant with respect to the pseudospin 
symmetry transformation
in RMF models with large scalar and vector potentials of nearly equal 
magnitude. In addition, the radial wave functions of this lower component 
are almost identical for two pseudospin partner states
\cite{Gin97,Gin98,Gin99a,Gin99b,Gin01a,Lev01,Gin02,Gin05b}.
The pseudospin symmetry was studied in various models
for nuclear structure
without and with deformation 
\cite{Men98,Men99,Sug00,Mar00,Alb01,Alb02,Sug02,Zho03,Gin04,Alb05,%
Ber06,Xu06,Lon06,Cas06}
and in nucleon-scattering 
\cite{Lee00,Gin01b,Lee04,Guo05b,Guo06}.
A symmetry breaking pseudospin-orbit potential 
can be derived from the wave equation of the
lower component. In principle, this potential should be a measure
for the strength of the symmetry breaking. It should appear as a small
perturbation in comparison with the dominating central potential
in a mean-field description of nuclei. However, this potential exhibits a
pole at finite radii and its position depends strongly 
on the level energy itself. Therefore, one cannot really estimate
the strength of the splitting and test how the choice of the various
contributions to the interaction influences the symmetry restoration
or breaking as observed experimentally.
The problems related to this peculiarity
have been discussed in the literature, see, e.g., ref.\ \cite{Mar00}.

Besides the single-particle levels that constitute pseudospin pairs
there are several states that are (pseudo-)unpaired in a nucleus.
There are all the $\{np_{1/2}\}_{so}=\{(n-1)\tilde{s}_{1/2}\}_{ps}$ 
states corresponding to $\tilde{l}=0$ 
irrespective of the principal quantum number $n$ and 
the states for given pseudospin $\tilde{l}>0$
with (pseudo) principal quantum number $n=1$ ($\tilde{n}=0$).
E.g., for the pseudoangular momentum $\tilde{l}=1$ 
the $1s_{1/2}$ level has no pseudospin partner
whereas for $\tilde{n} = n > 1$ the pairs 
$\{(n+1)s_{1/2},nd_{3/2}\}_{so}=
\{\tilde{n}\tilde{p}_{1/2},\tilde{n}\tilde{p}_{3/2}\}_{ps}$ 
appear. See the right part of figure \ref{fig:01}.
The latter feature is a typical observation for a system of levels
respecting exact supersymmetry. It is therefore tempting to apply 
supersymmetric quantum mechanics to the nuclear single-particle shell model
and to go beyond the harmonic oscillator case \cite{Cas92,Bal92}.
As will be shown below, the approach employing supersymmetry allows
to extract a pseudospin breaking potential that is
regular in contrast to the standard relativistic description.
From the condition of vanishing perturbation potential, general
potentials can be derived that exhibit exact pseudospin symmetry.
In order to compare the standard description of
pseudospin symmetry with the supersymmetric approach, the relativistic
Hamiltonian will be used as a starting point 
in the theoretical considerations below.
In principle, however, the supersymmetric approach can also be applied
to non-relativistic Hamiltonians.

The paper is organized as follows. In section \ref{sec:RMF}
the basic equations for describing single-particle states in 
relativistic models are given and the notation for quantum numbers,
wave functions, potentials and Hamiltonians is established.
The conventional description of pseudospin symmetry
in the relativistic approach is discussed in section \ref{sec:PS}
and a simple example is introduced that will accompany the 
following theoretical presentation.
In section \ref{sec:SUSY}, first the concept of supersymmetric 
quantum mechanics is presented as far as necessary 
to solve the problem considered here.
More detailed accounts 
can be found, e.g., in refs.\ \cite{Lah90,Coo95,Kal97,Coo01}.
Then the supermomenta and supersymmetric
partner Hamiltonians are defined. The form of the
reduced supermomenta at small radii 
leads to a particular choice of energy shifts
that relate the pseudospin partner 
Hamiltonians of the relativistic upper component
wave functions to the partner Hamiltonians in the supersymmetric
approach. Then, the connection between the wave functions in
the standard relativistic and the supersymmetric description
is given. A regular symmetry breaking 
potential is derived that quantifies the pseudospin splitting.
It can be used in the future to compare and
improve nuclear structure calculations.
In the final subsection, general potentials in relativistic Hamiltonians
with exact pseudospin symmetry are discussed.
The last section contains conclusions and an outlook.

\section{Single-particle states in relativistic mean-field models}
\label{sec:RMF}

The motion of the nucleons in RMF models is
governed by strong scalar and vector potentials 
(or self-energies) originating from the
minimal coupling of scalar and vector mesons to the nucleons. In addition, 
a tensor potential can appear if vector mesons are coupled non-minimally to
the meson field tensors, i.e.\ by a derivative coupling. 
In general, there are both isoscalar and isovector 
contributions to the nucleon self-energies.
The details of the particular RMF model are not important 
in the present discussion. See, e.g., references
\cite{Wal74,Ser86,Rei89,Rin96,Typ99,Hof01,Bue02,Ben03} 
for a more extensive presentation. Here, it suffices to consider only
the calculation of the single-particle states for given scalar, vector and
tensor potentials in the Dirac equation.

In the following spherical symmetry will be assumed. 
In this case, single-particle levels are characterized by
the principal quantum number $n=1,2,\dots$, 
the orbital angular momentum $l=0,1,\dots$,
the parity $\pi=(-1)^{l}$,
the total angular momentum $j=l\pm 1/2$,
and its projection $\Omega = -j , -j+1, \dots j-1, j$. 
In the relativistic description it is
convenient to combine the quantum numbers $l$ and $j$ by introducing the
quantity $\kappa=\pm 1, \pm 2, \pm 3, \dots$ where
$j(\kappa)=|\kappa|-1/2$ and
\begin{equation}
 l(\kappa) = \left\{ \begin{array}{ccl}
 \kappa -1 & \mbox{if} & \kappa > 0  \\
 -\kappa & \mbox{if} & \kappa < 0 \end{array} \right. \: .
\end{equation}
See table \ref{tab:kjl} for the relation between the various
quantum numbers.
Levels with different $\Omega$ for given $n$ and $\kappa$ are
degenerate in energy.


A single-particle state $\psi_{n\kappa\Omega}$ with energy $E_{n\kappa}$ 
(including the rest mass $m$) of a nucleon is
obtained by solving the time-independent Dirac equation 
\begin{equation}
 \mathcal{H} \psi_{n\kappa\Omega} = E_{n\kappa} \psi_{n\kappa\Omega}
\end{equation}
where
\begin{equation}
 \label{eq:H_Dirac}
 \mathcal{H} = \vec{\alpha} \cdot \vec{p} + V(r) 
 + \beta \left[ m - S(r) \right]
 - i \beta \vec{\alpha} \cdot \vec{T}(r)
\end{equation}
is the relativistic Hamiltonian
with momentum operator $\vec{p}$ and 
Dirac matrices $\vec{\alpha}$ and $\beta$ in standard notation.
The system of units is chosen such that $\hbar = c = 1$.
The scalar, vector and tensor potentials are denoted by $S(r)$, $V(r)$ and
$\vec{T}(r) = T(r) \vec{r}/r$. Here, the sign convention with a positive scalar
field $S$ is used. 
For spherical symmetry it is customary to write the Dirac spinor as
\begin{equation}
 \label{eq:psi_Dirac}
 \psi_{n\kappa\Omega} = \frac{1}{r} \left( \begin{array}{c} 
 F_{n\kappa}(r) \mathcal{Y}_{\kappa \Omega}(\hat{r})  \\ 
 i G_{n\kappa}(r) \mathcal{Y}_{-\kappa \Omega}(\hat{r}) \end{array} \right)
\end{equation}
with radial wave functions $F_{n\kappa}(r)= F_{nl_{j}}(r)$ 
and $G_{n\kappa}(r)=G_{nl_{j}}(r)$
in the upper and lower components, respectively, that
are normalized according to
\begin{equation}
 \int_{0}^{\infty} dr \: \left[ \left| F_{n\kappa}(r) \right|^{2}
 + \left| G_{n\kappa}(r) \right|^{2} \right] = 1 \: .
\end{equation}
The spin and angular dependence is contained in the 
spinor spherical harmonics
\begin{equation}
 \mathcal{Y}_{\kappa \Omega}(\hat{r})
 = \sum_{m_{l}m_{s}} (l \: m_{l} \: s \: m_{s} | j \: \Omega )
 Y_{lm_{l}}(\hat{r}) \chi_{sm_{s}}
\end{equation}
where the orbital angular momentum $l(\kappa)$ of the
spherical harmonic $Y_{lm_{l}}$ is coupled with the nucleon spin $s=1/2$ 
of the spinor $\chi_{sm_{s}}$ to the total angular momentum $j(\kappa)$.
The upper and lower component
in (\ref{eq:psi_Dirac})
have an opposite sign for the quantum number $\kappa$ in the
spinor spherical harmonics and consequently opposite parity with
a difference of one in the orbital angular momentum.
By applying the Hamiltonian (\ref{eq:H_Dirac}) to the Dirac spinor
(\ref{eq:psi_Dirac}) and using the relations
\begin{equation}
 \label{eq:sigmadotr}
 \vec{\sigma} \cdot \frac{\vec{r}}{r} \mathcal{Y}_{\kappa \Omega}(\hat{r})
 = -  \mathcal{Y}_{-\kappa \Omega}(\hat{r}) 
\end{equation}
and
\begin{equation}
 \vec{\sigma} \cdot \vec{p} \frac{f(r)}{r} \mathcal{Y}_{\kappa \Omega}(\hat{r})
 = \frac{i}{r} \left( \frac{d}{dr} - \frac{\kappa}{r} \right) f(r)
 \mathcal{Y}_{-\kappa \Omega}(\hat{r}) \: ,
\end{equation}
the set of coupled first-order differential equations
\begin{eqnarray}
 \label{eq:cplG}
 \left( E_{n\kappa}-V+m-S \right) G_{n\kappa}(r) & = & 
 \left( \frac{d}{dr} - \frac{\kappa}{r} - T \right) F_{n\kappa}(r)
 \\ 
  \left( E_{n\kappa}-V-m+S \right) F_{n\kappa}(r) & = & -  
\left( \frac{d}{dr} + \frac{\kappa}{r} + T \right) G_{n\kappa}(r)
\end{eqnarray}
for the radial wave functions is obtained.
A simple decoupling leads to the independent wave equations
\begin{eqnarray}
 \label{eq:HF}
 H_{F}(\kappa) F_{n\kappa} & = & E_{n\kappa} F_{n\kappa}
 \\
 \label{eq:HG}
 H_{G}(\kappa) G_{n\kappa} & = & E_{n\kappa} G_{n\kappa}
\end{eqnarray}
with Schr\"{o}dinger-like Hamiltonians
\begin{eqnarray}
 \label{eq:HF2}
 H_{F}(\kappa) & = &
 - \left( \frac{d}{dr} + \frac{\kappa}{r} + T \right) 
 \frac{1}{A} 
 \left( \frac{d}{dr} - \frac{\kappa}{r} - T \right) + V- S + m
 \\ \nonumber & = &
 \frac{1}{A} \left[
  - \frac{d^{2}}{dr^{2}} + \frac{\kappa(\kappa-1)}{r^{2}}
 + 2T \frac{\kappa}{r} + T^{2}
 + \frac{A^{\prime}}{A} 
 \left( \frac{d}{dr} - \frac{\kappa}{r} - T \right) \right]
 \\ \nonumber & & 
  + V - S + m
\end{eqnarray}
and
\begin{eqnarray}
 \label{eq:HG2}
 H_{G}(\kappa) & = & \frac{1}{A}
 \left[ - \frac{d^{2}}{dr^{2}} + \frac{\kappa(\kappa+1)}{r^{2}}
 + 2T\frac{\kappa}{r} + T^{2}
 + \frac{B^{\prime}}{B}\left( \frac{d}{dr} + \frac{\kappa}{r} + T\right) 
 \right]
 \\ \nonumber & & 
  + V - S + m \: ,
\end{eqnarray}
respectively, that explicitly depend on the quantum number $\kappa$. 
A prime denotes the derivative with respect to the radius $r$.
The quantities
\begin{equation}
 A(r) = E_{n\kappa}-V(r)+m-S(r)
\end{equation}
and
\begin{equation}
 \label{eq:defB}
 B(r) = E_{n\kappa}-V(r)-m+S(r)
\end{equation}
depend on the single-particles energy $E_{n\kappa}=E_{nl_{j}}$ 
appearing as the 
eigenvalue in equations (\ref{eq:HF}) and (\ref{eq:HG}).
The Hamiltonians $H_{F}(\kappa)$ and $H_{G}(\kappa)$ have the same
spectrum, even though the potentials have a different form.
$A(r)$ can be considered as twice the position depending 
effective nucleon mass $m_{\rm eff}$. 
It is always positive in standard RMF models.
The dependence on the energy $E_{n\kappa}$ is a relativistic effect.
Since $E_{n\kappa}$ contains the rest mass $m$ of the nucleon, the variation
of $A(r)$ with a change in the nucleon binding energy is very small.
For vanishing scalar and vector fields $S(r)$ and $V(r)$ at large radii $r$, 
$B(r)$ reduces to the non-relativistic energy 
$\varepsilon_{n\kappa} = E_{n\kappa}-m<0$ of the bound nucleon.
The Hamiltonian (\ref{eq:HF}) 
for the upper component wave function $F_{n\kappa}(r)$ can be written 
in the familiar non-relativistic form
\begin{equation}
 \label{eq:HFnr}
 H_{F}(\kappa) =
  - \frac{d}{dr}\frac{1}{2m_{\rm eff}}\frac{d}{dr} 
 + \frac{l(l+1)}{2m_{\rm eff}r^{2}}
 + V_{c}(r) + V_{so}(r) \langle \vec{l} \cdot \vec{s} \rangle + m
\end{equation}
with central and spin-orbit potentials
\begin{eqnarray}
 \label{eq:Vc}
 V_{c}(r) & = & V - S
 + \frac{T}{A} \left( T - \frac{A^{\prime}}{A} \right)
 + \frac{1}{Ar} \left( 2T - \frac{A^{\prime}}{A} \right)
 \\ \nonumber
 & = & V - S - \frac{T^{2}}{2m_{\rm eff}}
 + \frac{1+Tr}{2} V_{so}(r)
 \\
 \label{eq:Vso}
 V_{so}(r) & = & \frac{2}{Ar} \left( 2 T - \frac{A^{\prime}}{A} \right)
 = \frac{1}{m_{\rm eff}r} 
\left[ 2 T + \frac{1}{2m_{\rm eff}} \frac{d}{dr} \left(V+S\right)\right]
\end{eqnarray}
since $\kappa(\kappa-1) = l(l+1)$ and $\kappa-1 = 2 \langle \vec{l} \cdot
\vec{s} \rangle$.  The spin-orbit potential contains contributions from the
tensor potential $T$ and the derivative of the sum $V+S$ whereas the main
contribution to the central potential arises from the difference $V-S$.
With the typical strong vector and scalar potentials 
of almost similar magnitude
inside nuclei, the usual depth 
of the central potential and the strong 
surface-peaked spin-orbit potential are nicely explained in the
relativistic approach. A tensor potential is usually omitted in
traditional RMF calculations but it has a distinct effect
on the spin-orbit splitting that can be relevant for
the size of the shell gaps.

\section{Relativistic description of pseudospin symmetry}
\label{sec:PS}

The basic idea in the relativistic description of pseudospin symmetry
is a comparison of the Hamiltonians $H_{G}(\kappa)$, $H_{G}(\kappa^{\prime})$
and of the radial wave functions 
$G_{(n+1)\kappa}(r)$, $G_{n\kappa^{\prime}}(r)$
for the two pseudospin partner levels of the doublet
$\{(n+1)\kappa,n\kappa^{\prime}\}_{so}$. 
It is easy to see that for every
state with $\kappa > 0$ the pseudospin partner state has the
quantum number $\kappa^{\prime}=-\kappa-1$. 
The corresponding orbital angular momenta
are $l=\kappa-1$ and $l^{\prime}=-(-\kappa-1)=\kappa+1$, respectively. 
Similarly, the
total angular momenta are given by $j=\kappa-1/2$ and
$j^{\prime}=|-\kappa-1|-1/2=\kappa+1/2$, respectively. 
Hence, the relation between
$\tilde{l}$ and $\kappa$ is given by
\begin{equation}
 \tilde{l}(\kappa) = \left\{ \begin{array}{ccl}
 \kappa  & \mbox{if} & \kappa > 0  \\
 -\kappa-1 & \mbox{if} & \kappa < 0 \end{array} \right. \: .
\end{equation}
The single-particle states
in the relativistic model with $\kappa=-1$, i.e.\ the $p_{1/2}$
states, have no pseudospin partner. In this case $\tilde{l} = 0$.
See table \ref{tab:kjl} for a summary of the relations for the
quantum numbers.

\begin{figure}[t]
\center{\includegraphics[width=12cm,angle=0]{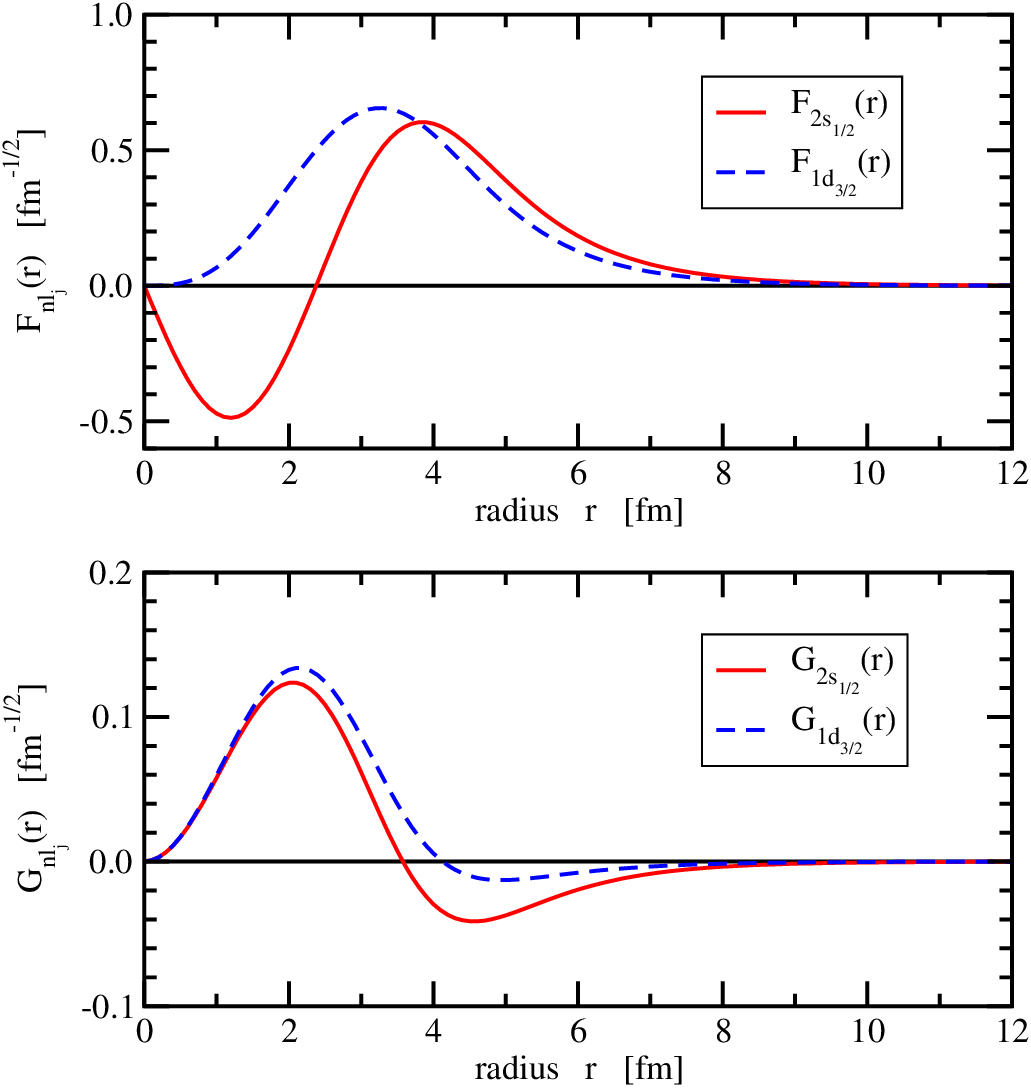}}
\caption{\label{fig:02}
Wave functions of the upper (top panel) and lower (bottom panel) component
in the Dirac spinor 
for the $2s_{1/2}$ (red solid line) and $1d_{3/2}$ (blue dashed line)
pseudospin partner levels.}
\end{figure}

A simple example shows the main features in the relativistic description
of pseudospin partner states. Here, the lowest pair of levels with
$\tilde{l}=1$,
i.e., $\{2s_{1/2},1d_{3/2}\}_{so},$ will be considered. 
The shape of scalar and vector potentials
in RMF models is often well approximated by a Woods-Saxon form, i.e.,
$S(r) = S_{0} f(r,R,a)$ and $V(r) = V_{0} f(r,R,a)$ with
$f(r,R,a) = 1/\{1+\exp[(r-R)/a]\}$. Assuming a radius of $R=3.8$~fm,
a diffuseness of $a=0.65$~fm and absolute magnitudes of
$S_{0} = 450$~MeV and $V_{0} = 370$~MeV, respectively, (corresponding to the
the mean-field potentials of neutrons in a nucleus like ${}^{40}$Ca) one finds
single-particles energies of $E_{2s_{1/2}}-m=-15.604$~MeV 
and $E_{1d_{3/2}}-m=-15.424$~MeV with
the $1d_{3/2}$ level lying only $0.180$~MeV above the $2s_{1/2}$ level.
The corresponding wave functions of the upper and of the lower
component in the Dirac spinor
are depicted in the top and bottom panel of figure \ref{fig:02}.
The wave functions $F_{2s_{1/2}}(r)$ and $F_{1d_{3/2}}(r)$ are very different
where the former has an additional node because there exists
a stronger bound $1s_{1/2}$ state at $E_{1s_{1/2}}-m=-56.848$~MeV 
without node (except at $r=0$). 
In contrast, the wave functions
$G_{2s_{1/2}}(r)$ and $G_{1d_{3/2}}(r)$ resemble each other closely, a sign
for the almost perfect pseudospin degeneracy of the two states.
Due to the different orbital angular momenta and corresponding centrifugal
potentials in $H_{F}(\kappa)$ and $H_{F}(\kappa^{\prime})$ there is
a clear difference in the radial dependence of the 
upper component wave functions at small $r$. But for the 
lower component radial wave function the $r$ dependence is very similar.


In fact, the orbital angular momenta in the Hamiltonians $H_{G}(\kappa)$
and $H_{G}(-\kappa-1)$ of a pseudospin pair 
$[(n+1)\kappa,n(-\kappa-1)]_{so}$
for $\kappa = \tilde{l} >0$
are identical since $l(-\kappa) = \kappa = l(\kappa+1)$. The centrifugal
potential $\kappa(\kappa+1)/(Ar^{2})$ in (\ref{eq:HG2}) is invariant
with respect to the replacement $\kappa \to -\kappa-1$. The difference
between the potentials
\begin{equation}
 \label{eq:delta1}
 \Delta(\tilde{l},r) = H_{G}(\kappa) - H_{G}(-\kappa-1)
 = \frac{2\kappa+1}{Ar} \left( \frac{B^{\prime}}{B} + 2 T \right)
\end{equation}
could be considered as a measure for the pseudospin splitting
because for $\Delta(\tilde{l})=0$ the Hamiltonians $H_{G}(\kappa)$ and
$H_{G}(-\kappa-1)$ are identical and yield the same energy
eigenvalues. 

\begin{figure}[t]
\center{\includegraphics[width=12cm,angle=0]{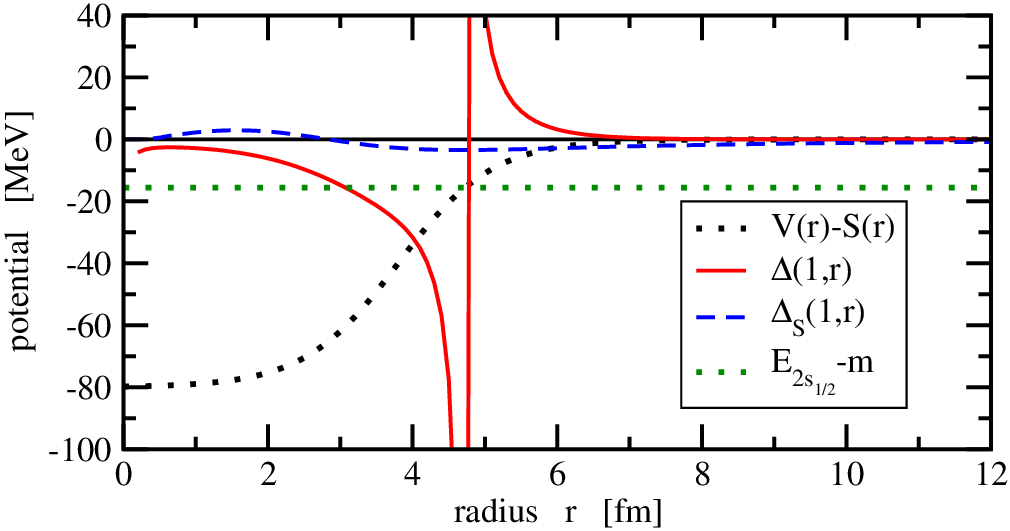}}
\caption{\label{fig:03}
Pseudospin symmetry breaking potentials in the standard relativistic
description (red solid line) and the supersymmetric approach (blue dashed
line). The potential difference $V(r)-S(r)$ is denoted by a black dotted line
and the energy of the $2s_{1/2}$ state by a horizontal green dotted line.}
\end{figure}

However, there is a problem in this approach 
since the quantity $B(r)$ has a zero at finite $r$ inside
the nucleus with usual scalar and vector potentials in RMF models. 
This is most easily seen in figure \ref{fig:03} where the potential
difference $V(r)-S(r)$ and the non-relativistic energy $E_{2s_{1/2}}-m$
are plotted. They are identical at a certain radius $r$ and the quantity
(\ref{eq:defB}) is zero. Consequently, the difference potential 
(\ref{eq:delta1}), shown as a red solid line in figure \ref{fig:03},
has a pole. Its position is strongly energy dependent. 
Thus, $\Delta(\tilde{l})$ 
cannot be considered small as compared to the
potential $V-S$ in the differential equation 
and it is therefore not a good measure for the symmetry breaking.

For vanishing tensor potential $T=0$, the pseudospin-orbit potential
(\ref{eq:delta1}) is zero 
for $B^{\prime}=0$ or $S(r) = V(r) + C$ with a constant $C$.
As a consequence of this condition,
the main contribution $V-S$ to the central potential (\ref{eq:Vc})
is just an overall shift of the energy scale for exact pseudospin symmetry.
But then there will be no sufficiently strong 
central potential to bind the states as required in a nucleus. 

A case where exact pseudospin symmetry is found is the relativistic
harmonic oscillator with $V(r) = -S(r) = m\omega^{2}r^{2}/4$ and $T=0$
\cite{Cas06,Mot03,Lis04,Gin05a,Guo05a}.
The effective mass $m_{\rm eff} = A/2=(E+m)/2$ is a constant,
$B(r) = E-m-m\omega^{2}r^{2}/2$ but $B^{\prime}(r) \neq 0$ 
and $\Delta(\tilde{l},r) \neq 0$.
Hence, the identification of $\Delta(\tilde{l},r)$ as the relevant 
quantity to measure the pseudospin splitting is doubtful.
It would be convenient to have an independent measure of the 
symmetry breaking
that has a regular behaviour for all radii
and that vanishes for exact pseudospin symmetry.

\section{Supersymmetric description of pseudospin symmetry}
\label{sec:SUSY}

The problems related to the conventional relativistic explanation
of pseudospin symmetry call for an alternative approach that
allows to derive a regular symmetry breaking potential which is a small 
perturbation without a pole and useful for a quantitative comparison of 
different models.
The main idea of the supersymmetric approach to describe the pseudospin
symmetry in nuclei is similar to the standard relativistic description.
In both cases one does not compare the original model Hamiltonians of the
pseudospin partner systems but closely related Hamilton operators with the
same spectra.
In fact, it will be shown below that a comparison of the supersymmetric partner
Hamiltonians provides a way to define a proper pseudospin splitting potential.
These partner potentials are derived from the Hamiltonians 
$H_{F}(\kappa)$ and $H_{F}(\kappa^{\prime})$ for the upper component 
wave functions of the pseudospin partners $\kappa$ and 
$\kappa^{\prime}=-\kappa-1$, respectively. To this end, the Hamiltonian
(\ref{eq:HF2}) is written as
\begin{equation}
 \label{eq:HFH1}
 H_{F}(\kappa) = H_{1}(\kappa) + E(\kappa)
\end{equation}
with a Hamiltonian $H_{1}(\kappa)$ of a supersymmetric pair and
an energy shift $E(\kappa)$ to be determined below.
A corresponding relation holds for the pseudospin partner Hamiltonian
$H_{F}(\kappa^{\prime})$.

\subsection{General features of supersymmetric quantum mechanics}

It is well known, that
every second-order Hamiltonian can be factorized in a product
of two Hermitian conjugate first-order operators \cite{Sch40,Inf41,Inf51}. 
In the present case, operators $B_{\kappa}^{+}$ and
$B_{\kappa}^{-} = \left[ B_{\kappa}^{+} \right]^{\dagger}$ are introduced
such that
\begin{equation}
 \label{eq:H1def}
 H_{1}(\kappa) = B^{+}_{\kappa}B^{-}_{\kappa}  \: .
\end{equation}
Then, the Hermitian operators 
\begin{equation}
 \label{eq:supercharge}
 Q_{1}(\kappa) = \left( \begin{array}{cc}
 0 & B_{\kappa}^{+} \\ B_{\kappa}^{-} & 0
 \end{array} \right)
 \qquad
  Q_{2}(\kappa) = iQ_{1}(\kappa) \tau = \left( \begin{array}{cc}
 0 & -iB_{\kappa}^{+} \\ iB_{\kappa}^{-} & 0
 \end{array} \right)
\end{equation}
are formed that are so-called supercharges with respect
to the involution
\begin{equation}
 \tau = \tau^{\dagger} = \left( \begin{array}{cc}
 1 & 0 \\ 0 & -1
 \end{array} \right)
 \qquad
 \tau^{2} = 1
\end{equation}
because $\{Q_{1},\tau \} = \{Q_{2},\tau \} = 0$.
In the next step, the supersymmetric Hamiltonian
\begin{equation}
 \label{eq:HS}
 H_{S}(\kappa) = \left[Q_{1}(\kappa)\right]^{2}
 = \left[Q_{2}(\kappa)\right]^{2}
 =  \left( \begin{array}{cc}
 H_{1}(\kappa) & 0  \\ 0 & H_{2}(\kappa)
 \end{array} \right)
\end{equation}
is obtained with the supersymmetric partner Hamiltonians (\ref{eq:H1def}) and
\begin{equation}
 \label{eq:H2}
 H_{2}(\kappa) = B^{-}_{\kappa}B^{+}_{\kappa}  
\end{equation}
on the diagonal.
Obviously, $H_{S}(\kappa)$ commutes with $Q_{1}(\kappa)$ and $Q_{2}(\kappa)$, 
i.e.,
\begin{equation}
 \label{eq:alg1}
 \left[ H_{S}(\kappa), Q_{1}(\kappa) \right] =
 \left[ H_{S}(\kappa), Q_{2}(\kappa) \right] = 0 
\end{equation}
and the supercharges $Q_{i}(\kappa)$ ($i=1,2$) are
generators of the symmetry transformation.
The supercharges (\ref{eq:supercharge}) and the Hamiltonian (\ref{eq:HS})
with the commutators (\ref{eq:alg1})
and the anticommutator
\begin{equation}
 \{ Q_{1}(\kappa), Q_{2}(\kappa)\} = 0
\end{equation}
are the most simple example of a supersymmetric algebra.
Since $H_{S}(\kappa)$ is the square of the Hermitian operators $Q_{i}(\kappa)$,
all eigenvalues $E_{S}(\bar{n}\kappa)$ of the eigenvalue equation
\begin{equation}
 \label{eq:HSev}
 H_{S}(\kappa) \Psi_{S}(\bar{n}\kappa) 
 = E_{S}(\bar{n}\kappa) \Psi_{S}(\bar{n}\kappa)
\end{equation}
with the two-component wave function
\begin{equation}
 \label{eq:psi_SUSY}
 \Psi_{S}(\bar{n}\kappa) = \left( \begin{array}{c} 
 \psi_{1}(\bar{n}\kappa) \\ \psi_{2}(\bar{n}\kappa)
 \end{array} \right)
\end{equation}
are non-negative. It is easily seen by applying $Q_{i}(\kappa)$ to equation
(\ref{eq:HSev}), that $H_{1}(\kappa)$ and $H_{2}(\kappa)$ 
have the same spectrum of
positive energies $E_{S}(\bar{n}\kappa)>0$ where the operators 
$B_{\kappa}^{+}$ and $B_{\kappa}^{-}$ connect the components
of the wave function (\ref{eq:psi_SUSY}) by the transformations
\begin{equation}
 \label{eq:Bpsi}
 \psi_{2}(\bar{n}\kappa) 
 = \frac{B_{\kappa}^{-}}{\sqrt{E_{S}(\bar{n}\kappa)}} \psi_{1}(\bar{n}\kappa)
 \qquad
 \psi_{1}(\bar{n}\kappa) 
 = \frac{B_{\kappa}^{+}}{\sqrt{E_{S}(\bar{n}\kappa)}} \psi_{2}(\bar{n}\kappa)
 \: .
\end{equation}
If there is an eigenstate $\Psi_{S}(0\kappa)$ ($\bar{n}=0$) with energy 
$E_{S}(0\kappa)=0$, the supersymmetry is called exact because 
$Q_{i}(\kappa) \Psi_{S}(0\kappa)=0$ and this ground state obeys the symmetry of
the Hamiltonian $H_{S}(\kappa)$.
In this case, 
\begin{equation}
 \label{eq:Bkm0}
 B_{\kappa}^{-}\psi_{1}(0\kappa)=0 \qquad \psi_{2}(0\kappa)=0
\end{equation}
or 
\begin{equation}
 B_{\kappa}^{+}\psi_{2}(0\kappa)=0 \qquad \psi_{1}(0\kappa)=0 \: ,
\end{equation}
i.e.,  the Hamiltonian $H_{1}(\kappa)$ or $H_{2}(\kappa)$ 
has an additional state at zero energy that is not appearing 
for its supersymmetric partner Hamiltonian. In this paper, the 
usual convention is chosen such that $H_{1}(\kappa)$ has a ground state
at zero energy if the supersymmetry is exact.
If there is no state $\Psi_{S}(0\kappa)$ at energy $E_{S}(0\kappa)=0$, 
the supersymmetry
is called broken since $Q_{i}(\kappa) \Psi_{S}(\bar{n}\kappa) \neq 0$
for all $\bar{n}=1,2,\dots$ and
the partner Hamiltonians $H_{1}(\kappa)$ and $H_{2}(\kappa)$ have identical
spectra.

\subsection{The supermomentum and supersymmetric partner Hamiltonians}

The main task in the supersymmetric description of pseudospin symmetry
is the construction of the operators $B_{\kappa}^{+}$ and $B_{\kappa}^{-}$.
The particular form of the Hamiltonian (\ref{eq:HF2}) suggests to
use the ansatz
\begin{equation}
 B_{\kappa}^{-}
 =  \frac{1}{\sqrt{A_{\kappa}(r)}} \left[ Q_{\kappa}(r)
 +  \frac{d}{dr} \right]
 \qquad
 B_{\kappa}^{+} 
 = \left[ Q_{\kappa} (r)
   - \frac{d}{dr} \right] \frac{1}{\sqrt{A_{\kappa}(r)}}
\end{equation}
with (the principal quantum number $n$ is suppressed in the following
if not explicitly needed)
\begin{equation}
 \label{eq:adef}
 A_{\kappa}(r) = E_{\kappa} - V(r) + m - S(r)
\end{equation}
and a quantity $Q_{\kappa}(r)$ that will be called supermomentum
by dimensional reasons. The traditional superpotential
is given by 
\begin{equation}
 \label{eq:Wk}
 W_{\kappa}(r) = \sqrt{\frac{2}{A_{\kappa}(r)}}Q_{\kappa}(r) \: ,
\end{equation}
however, it has the dimension
energy$^{1/2}$. The supersymmetric partner Hamiltonians are found
as
\begin{equation}
 \label{eq:H1}
 H_{1}(\kappa) = B^{+}_{\kappa}B^{-}_{\kappa}  =  
 \frac{1}{A_{\kappa}} \left[ Q_{\kappa}^{2} - Q_{\kappa}^{\prime} 
 - \frac{d^{2}}{dr^{2}}
 + \frac{A_{\kappa}^{\prime}}{A_{\kappa}}
 \left( Q_{\kappa} + \frac{d}{dr} \right) \right]
\end{equation}
and
\begin{equation}
 H_{2}(\kappa) = B^{-}_{\kappa}B^{+}_{\kappa} 
= \frac{1}{A_{\kappa}} \left[ Q_{\kappa}^{2} + Q_{\kappa}^{\prime} 
 - \frac{d^{2}}{dr^{2}} + \frac{A_{\kappa}^{\prime}}{A_{\kappa}} \frac{d}{dr}
  + \frac{A_{\kappa}^{\prime\prime}}{2A_{\kappa}}
 - \frac{3(A_{\kappa}^{\prime})^{2}}{4A_{\kappa}^{2}}\right] 
\end{equation}
with the symmetry
\begin{equation}
 \label{eq:Psym}
 H_{1}(\kappa,Q_{\kappa}) = 
 H_{2}(\kappa,-Q_{\kappa}-\frac{A_{\kappa}^{\prime}}{2A_{\kappa}}) \: .
\end{equation}
If the supersymmetry is exact, the supermomentum $Q_{\kappa}$
can be found from the ground state wave function $\psi_{1}(0\kappa)$ of the 
Hamiltonian $H_{1}(\kappa)$ as
\begin{equation}
 Q_{\kappa}(r) = - \frac{d}{dr} \ln \psi_{1}(0\kappa) 
\end{equation}
because of equation (\ref{eq:Bkm0}), i.e., it is given by the negative
logarithmic derivative of the groundstate wave function. 
Comparing equation (\ref{eq:H1}) via (\ref{eq:HFH1}) 
with equation (\ref{eq:HF}) 
leads to the defining equation for the supermomentum
$Q_{\kappa}(r)$ in general. 
However, it is useful to split off the dependence
on $\kappa$ and the tensor potential $T$ first and to introduce the
reduced supermomentum 
\begin{equation}
 \label{eq:Pp}
 q_{\kappa}(r) = Q_{\kappa}(r) + \frac{\kappa}{r} + T(r) 
\end{equation}
with the results
\begin{eqnarray}
 H_{1}(\kappa) & = &
  \frac{1}{A_{\kappa}} \left[ 
 - \frac{d^{2}}{dr^{2}} + \frac{\kappa(\kappa-1)}{r^{2}} 
 + q_{\kappa}^{2} - 2 q_{\kappa} \frac{\kappa}{r} - q_{\kappa}^{\prime}
 - 2 q_{\kappa} T + T^{2} 
 \right. \\ \nonumber & & \left.
 + 2 T \frac{\kappa}{r} + T^{\prime}
 + \frac{A_{\kappa}^{\prime}}{A_{\kappa}}
 \left( q_{\kappa} + \frac{d}{dr} - \frac{\kappa}{r} - T \right) \right]
\end{eqnarray}
and
\begin{eqnarray}
 \label{eq:H2p}
 H_{2}(\kappa) & = &
 \frac{1}{A_{\kappa}} \left[  
 - \frac{d^{2}}{dr^{2}} + \frac{\kappa(\kappa+1)}{r^{2}} 
 + q_{\kappa}^{2} - 2 q_{\kappa} \frac{\kappa}{r} + q_{\kappa}^{\prime}
 - 2 q_{\kappa} T  + T^{2}  
 \right. \\ \nonumber & & \left.
 + 2 T \frac{\kappa}{r} - T^{\prime}
 + \frac{A_{\kappa}^{\prime}}{A_{\kappa}} \frac{d}{dr}
  + \frac{A_{\kappa}^{\prime\prime}}{2A_{\kappa}}
 - \frac{3(A_{\kappa}^{\prime})^{2}}{4A_{\kappa}^{2}}\right] \: .
\end{eqnarray}
These two Hamiltonians show the symmetry
\begin{equation}
 H_{1}(\kappa,q_{\kappa},T) 
 = H_{2}(-\kappa,-q_{\kappa}-\frac{A_{\kappa}^{\prime}}{2A_{\kappa}},-T) \: ,
\end{equation}
however, it is irrelevant for the present considerations.
Subsequently, the defining equation for the reduced supermomentum 
\begin{equation}
 \label{eq:Ric}
  q_{\kappa}^{2} -  \left( 2 \frac{\kappa}{r} + 2 T -
 \frac{A_{\kappa}^{\prime}}{A_{\kappa}} \right) 
 q_{\kappa}  - q_{\kappa}^{\prime}
=  - A_{\kappa} B_{\kappa} - T^{\prime}
\end{equation}
with
\begin{equation}
 B_{\kappa}(r) =  E(\kappa) - V(r) + S(r) - m 
\end{equation}
is obtained. Note that $B_{\kappa}(r)$ depends on the energy shift $E(\kappa)$
and $A_{\kappa}(r)$ on the eigenvalue $E_{\kappa}$.
Equation (\ref{eq:Ric}) 
is the central relation that connects the reduced supermomentum $q_{\kappa}(r)$
with the relativistic potentials $S(r)$, $V(r)$ and $T(r)$.
In general, the reduced supermomentum $q_{\kappa}(r)$ is not uniquely determined
by equation (\ref{eq:Ric}) and suitable boundary conditions have to be
specified. In the present case, it will be required that
$q_{\kappa}(r)=0$ for $r=0$.

It is convenient, e.g., in numerical applications,  
to transform the 
Riccati equation (\ref{eq:Ric}), an inhomogeneous nonlinear first-order
differential equation, into a linear second-order
differential equation to determine the reduced supermomentum $q_{\kappa}(r)$
for given scalar, vector and tensor potentials
$S(r)$, $V(r)$ and $T(r)$.
To this end, a new function $y_{\kappa}(r) \neq 0$ 
is introduced that is related via
\begin{equation}
 q_{\kappa} = -\frac{y_{\kappa}^{\prime}}{y_{\kappa}} 
\end{equation}
to the reduced supermomentum $q_{\kappa}(r)$. 
With
\begin{equation}
 q_{\kappa}^{\prime} = -\frac{y_{\kappa}^{\prime\prime}}{y_{\kappa}} 
 +\frac{(y_{\kappa}^{\prime})^{2}}{y_{\kappa}^{2}}
\end{equation}
one immediately obtains the homogeneous differential equation
\begin{equation}
 \label{eq:ykd}
  y_{\kappa}^{\prime\prime} + \left( 2 \frac{\kappa}{r} + 2 T -
 \frac{A_{\kappa}^{\prime}}{A_{\kappa}} \right) y_{\kappa}^{\prime}
 + \left(A_{\kappa} B_{\kappa} + T^{\prime}\right) y_{\kappa} = 0 
\end{equation}
that can be solved with standard techniques.


Using equation (\ref{eq:Ric}) the supersymmetric
partner Hamiltonian (\ref{eq:H2p}) assumes the form
\begin{eqnarray}
 \label{eq:H2pp}
 H_{2}(\kappa) & = &
 \frac{1}{A_{\kappa}} \left[  
 - \frac{d^{2}}{dr^{2}} + \frac{\kappa(\kappa+1)}{r^{2}} 
 + 2 q_{\kappa}^{\prime}
 - \frac{A_{\kappa}^{\prime}}{A_{\kappa}} \left( q_{k} - \frac{d}{dr} \right)
 \right. \\ \nonumber & & \left.
 + T^{2} + 2 T \frac{\kappa}{r}  - 2T^{\prime}
  + \frac{A_{\kappa}^{\prime\prime}}{2A_{\kappa}}
 - \frac{3(A_{\kappa}^{\prime})^{2}}{4A_{\kappa}^{2}}\right] 
 - B_{\kappa} \: .
\end{eqnarray}
It will be useful for the comparison with the Hamiltonian
$H_{2}(\kappa^{\prime})$ of the pseudospin partner system.

\subsection{Asymptotics of the supermomenta} 

The asymptotic behaviour of the reduced supermomentum
for $r \to 0$ and $r \to \infty$ is easily derived from 
equation (\ref{eq:Ric})
assuming regular $q_{\kappa}(r)$ and regular potentials 
$S(r)$, $V(r)$ and $T(r)$.
The reduced superpotential has to vanish
for $r \to 0$ otherwise a diverging potential would appear in the Hamiltonian
$H_{F}(\kappa)$. For small radii the reduced supermomentum is given by
\begin{equation}
 \label{eq:pr0}
 q_{\kappa}(r) =  
 \frac{A_{\kappa}(0)B_{\kappa}(0)+T^{\prime}(0)}{2\kappa+1} r + O(r^{2}) \: ,
\end{equation}
i.e., it increases almost linearly with the radius.
At large radii it becomes a constant
\begin{equation}
 \label{eq:pkinf}
 q_{\kappa}(r) =  \sqrt{(E_{\kappa}+m)[m-E(\kappa)]} + O(r^{-1}) 
\end{equation}
for potentials $S(r)$, $V(r)$ and $T(r)$ that approach zero 
at large radii $r$.
The energy shift $E(\kappa)$
with the condition $E(\kappa) < m$ has still to be determined.
The asymptotic form (\ref{eq:pkinf}) translates to a dependence
\begin{equation}
 y_{\kappa}(r) \to \exp \left\{ - \sqrt{(E_{\kappa}+m)[m-E(\kappa)]} \: r 
 \right\}
\end{equation}
of the auxiliary function $y_{\kappa}$ in equation (\ref{eq:ykd})
for $r \to \infty$. Starting at large radii and integrating inwards
a stable solution for $y_{\kappa}(r)$ is found.
Of course, if the relativistic potentials do not approach zero at large radii,
a different asymptotic behaviour of the reduced supermomentum will be obtained.

Considering the relation (\ref{eq:Pp}), the full supermomentum $Q_{\kappa}(r)$ 
approaches the same constant as $q_{\kappa}(r)$ for large $r$. In contrast,
the asymptotic behaviour of $Q_{\kappa}(r)$ for $r \to 0$ is determined
by the angular momentum term $\kappa/r$. For $\kappa>0$, one finds
$\lim_{r \to 0} Q_{\kappa}(r) = - \infty$ and 
$\lim_{r \to 0} Q_{\kappa^{\prime}}(r) = \infty$ for $\kappa^{\prime} =
- \kappa -1 <0$. The standard superpotential (\ref{eq:Wk}) shows
the same asymptotic limits as the full supermomentum. 
From general considerations of supersymmetric
quantum mechanics it is known, that the type of supersymmetry is determined
by the asymptotics of the superpotential. 
If there is a change of
sign in $W_{\kappa}(r)$ when comparing the limits $r\to 0$ with $r \to \infty$,
exact supersymmetry follows and a single non-degenerate
state at zero energy exists. 
Conversely, 
if the superpotential $W_{\kappa}(r)$ does not change the sign 
in the above limits the supersymmetry is broken and all eigenstates
are doubly degenerate with positive energy.

\subsection{Application of supersymmetry to pseudospin
levels and energy shifts}

\begin{figure}[t]
\center{\includegraphics[width=13cm,angle=0]{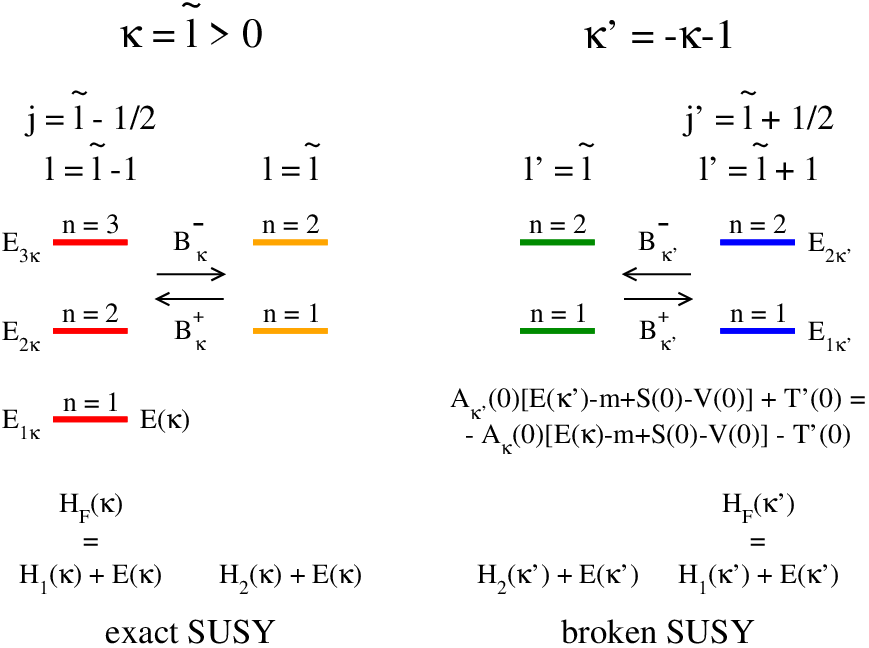}}
\caption{\label{fig:04}
Level scheme of pseudospin partner systems for given pseudoangular 
momentum $\tilde{l}$ in the supersymmetric description. 
See text for details.}
\end{figure}

With the foregoing considerations, the supersymmetric description can be 
applied to the system of pseudospin levels appearing in nuclei.
For given pseudo orbital angular momentum $\tilde{l} = \kappa \geq 1$
there is an unpaired state with total angular momentum 
$j = \tilde{l} - 1/2$, orbital angular momentum $l=\tilde{l}-1$ and
energy $E_{1\kappa}$, cf.\ the left part of figure \ref{fig:04}.
Choosing the energy shift as
\begin{equation}
 E(\kappa) = E_{1\kappa}
\end{equation}
the ground state energy of the Hamiltonian 
$H_{1}(\kappa)=H_{F}(\kappa)-E(\kappa)$ is placed at
zero energy and the case of exact supersymmetry is obtained.
The supersymmetric partner Hamiltonian $H_{2}(\kappa)$ has the same energy
eigenvalues as $H_{1}(\kappa)$ except for the zero-energy level.
The orbital angular momentum in $H_{2}(\kappa)$ is $l=\kappa=\tilde{l}$
as seen in the centrifugal potential in equation (\ref{eq:H2p}).

The pseudospin partner levels with $j^{\prime} = \tilde{l} + 1/2$ 
and $l^{\prime}=\tilde{l}+1$ are eigenstates
of the Hamiltonian $H_{F}(\kappa^{\prime})$ with
$\kappa^{\prime} = -\tilde{l} -1 <0$, see the right part of figure
\ref{fig:04}. The lowest state has a larger energy $E_{1\kappa^{\prime}}$
than the ground state of $H_{F}(\kappa)$. For this system
the case of broken supersymmetry applies where 
$H_{F}(\kappa^{\prime}) = H_{1}(\kappa^{\prime})+E(\kappa^{\prime})$
and the supersymmetric partner Hamiltonian
$H_{2}(\kappa^{\prime})+E(\kappa^{\prime})$
have identical eigenvalues. 
The orbital angular momentum in $H_{2}(\kappa^{\prime})$
is again given by $l^{\prime} = \kappa = \tilde{l}$ due to
$\kappa^{\prime}(\kappa^{\prime}+1) = (-\kappa-1)(-\kappa-1+1)
= \kappa(\kappa+1)$.
The energy shift $E(\kappa^{\prime})$ is now chosen such that
the asymptotic form of the reduced supermomentum $q_{\kappa^{\prime}}(r)$
is the same as for the reduced supermomentum $q_{\kappa}(r)$ in the limit
$r \to 0$. This condition
leads with equation (\ref{eq:pr0}) to the relation
\begin{equation}
 A_{\kappa}(0)B_{\kappa}(0)+T^{\prime}(0)
 = - \left[ A_{\kappa^{\prime}}(0)B_{\kappa^{\prime}}(0)+T^{\prime}(0) \right]
\end{equation}
since $2\kappa^{\prime}+1 = -(2\kappa+1)$. For exact pseudospin symmetry,
$E_{(n+1)\kappa} = E_{n\kappa^{\prime}}$ and consequently
$A_{\kappa}(r)=A_{\kappa^{\prime}}(r)$. Even if the
pseudospin symmetry is broken, $A_{\kappa}$ and $A_{\kappa^{\prime}}$
are almost identical and the difference can be neglected as 
a small relativistic correction. 
Without the contribution of the tensor potential
$T^{\prime}(0)$, the energy shift $E(\kappa^{\prime})$ is determined by
\begin{equation}
 E(\kappa^{\prime}) - m + S(0) - V(0) 
 = - \left[ E(\kappa) - m + S(0) - V(0) \right] \: .
\end{equation}

\begin{figure}[t]
\center{\includegraphics[width=12cm,angle=0]{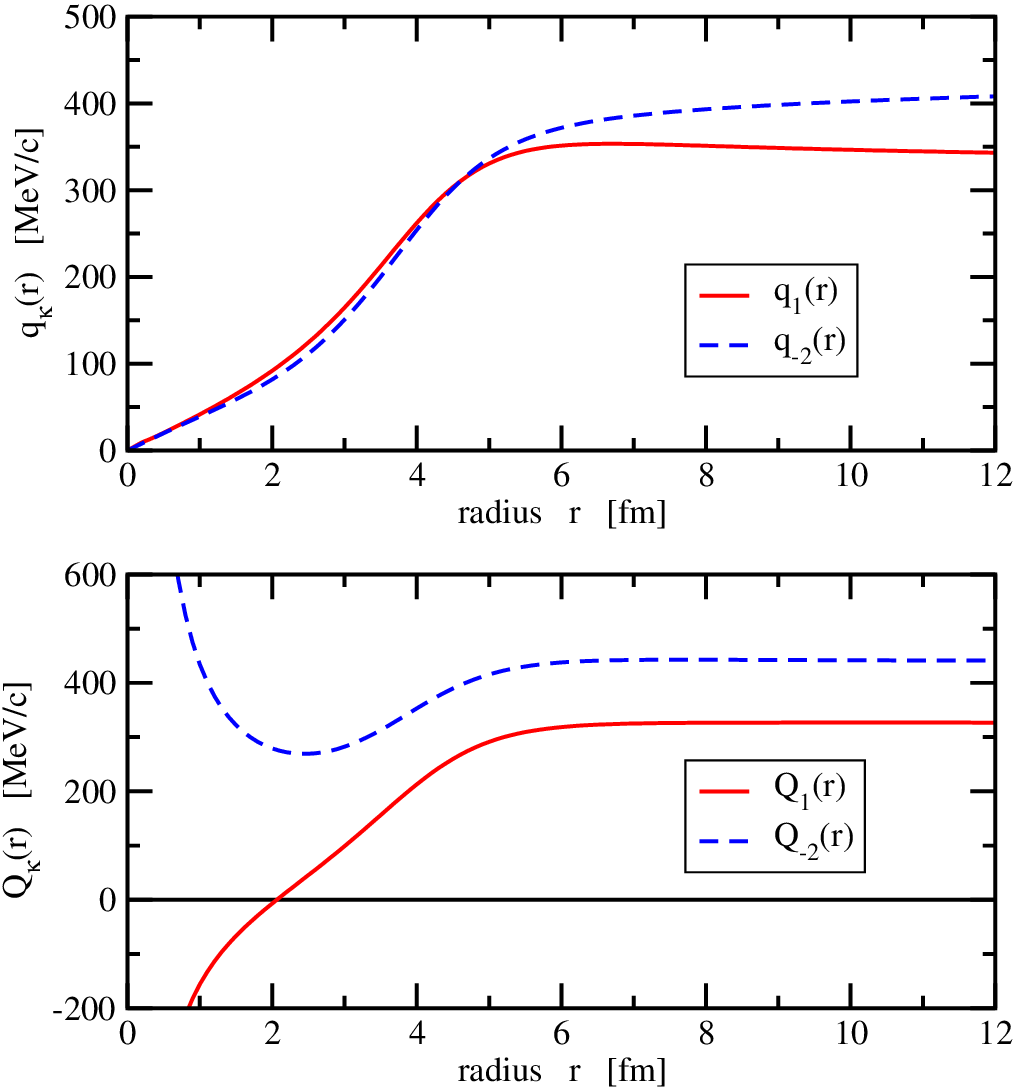}}
\caption{\label{fig:05}
Reduced supermomenta $q_{\kappa}(r)$ (top panel) 
and full supermomenta $Q_{\kappa}(r)=q_{\kappa}(r)-\kappa/r$ (bottom panel)
for the $2s_{1/2}$, $\kappa=1$ (red solid lines) and $1d_{3/2}$, 
$\kappa^{\prime} = -2$ (blue dashed lines) pseudospin partner levels.}
\end{figure}

In the top panel of figure \ref{fig:05} 
the reduced supermomenta $q_{\kappa}(r)$ 
and $q_{\kappa^{\prime}}(r)$
of the pseudospin pair levels with $\kappa = 1$ and $\kappa^{\prime}=-2$
are shown for the example in section \ref{sec:PS}. With the present choice
of the energy shifts $E(1)-m=-56.849$~MeV and
$E(-2)-m=-102.680$~MeV,
the reduced supermomenta display the same approximately
linear behaviour for $r \to 0$ as required. In fact, they are almost identical
for radii below 5~fm. At large $r$ they become constant and approach
different values for $r \to \infty$ as predicted by equation (\ref{eq:pkinf})
with $q_{1}(r) < q_{-2}(r)$ because $E(1) > E(-2)$. The full supermomenta
$Q_{1}(r)$ and $Q_{-2}(r)$ are shown in the bottom
panel of figure \ref{fig:05}. They contain the angular momentum contribution
and, hence, they diverge for $r \to 0$. The change of sign for
$Q_{1}(r)$ indicates the case of exact supersymmetry with a single state
at zero energy for $H_{1}(1)$. In contrast, there is no change of sign
for $Q_{-2}(r)$ corresponding to the case of broken supersymmetry.

\subsection{Relation of wave functions}

\begin{figure}[t]
\center{\includegraphics[width=12cm,angle=0]{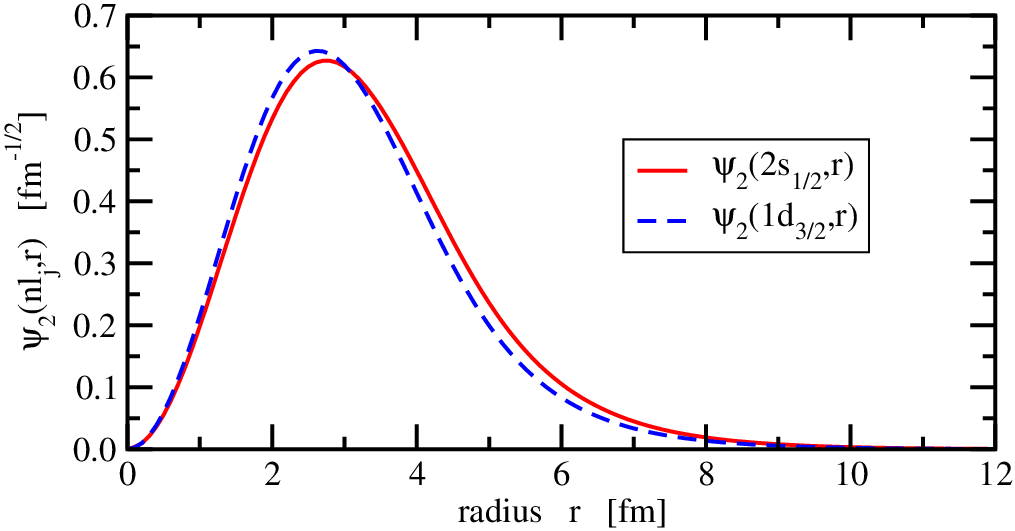}}
\caption{\label{fig:06}
Wave functions of the supersymmetric partner states
for the $2s_{1/2}$ (red solid line) and $1d_{3/2}$ (blue dashed line)
pseudospin partner levels.}
\end{figure}

The supersymmetric description of pseudospin degenerate states
allows to establish relations between the
wave functions in the Dirac spinor (\ref{eq:psi_Dirac}) and in
the supersymmetric state vector (\ref{eq:psi_SUSY}). According
to equation (\ref{eq:cplG}) the  lower component wave function of the spinor
is given by
\begin{equation}
 \label{eq:Gk2}
 G_{n\kappa}(r) = \frac{1}{A_{n\kappa}}  
 \left( \frac{d}{dr} - \frac{\kappa}{r} - T\right) F_{n\kappa}(r)
\end{equation}
for the pseudospin partner state $n\kappa$ with $\kappa>0$
and similarly for the state $n^{\prime}\kappa^{\prime}$ 
with $n^{\prime}=n-1$ and $\kappa^{\prime}=-\kappa-1$.
Using equation (\ref{eq:Bpsi}) the supersymmetric partner 
\begin{equation}
 \psi_{2}(n\kappa,r) = \frac{1}{\sqrt{[E_{n\kappa}-E(\kappa)]
 A_{n\kappa}}}
 \left( q_{n\kappa} - \frac{\kappa}{r} - T + \frac{d}{dr}\right)
 \psi_{1}(n\kappa,r)
\end{equation}
of the upper component wave function
$\psi_{1}(n\kappa,r)=F_{n\kappa}(r)$ is obtained
where the eigenenergy $E_{S}(\bar{n}\kappa)$ in the supersymmetric
formulation is identified with the energy 
$E_{n\kappa}-E(\kappa)=E_{n\kappa}-E_{0\kappa}$.
With equation (\ref{eq:Gk2}) one finds
\begin{equation}
 \label{eq:psi2}
 \psi_{2}(n\kappa,r) = 
 \sqrt{\frac{A_{n\kappa}}{E_{n\kappa}-E(\kappa)}} \left[
 \frac{q_{n\kappa}}{A_{n\kappa}} F_{n\kappa}(r)
 + G_{n\kappa}(r) \right]
\end{equation}
and
\begin{equation}
 \label{eq:psi2p}
 \psi_{2}(n^{\prime}\kappa^{\prime},r)=
 \sqrt{\frac{A_{n^{\prime}\kappa^{\prime}}}{E_{n^{\prime}\kappa^{\prime}}
 -E(\kappa^{\prime})}}  
 \left[ \frac{q_{n^{\prime}\kappa^{\prime}}}{A_{n^{\prime}\kappa^{\prime}}} 
  F_{n^{\prime}\kappa^{\prime}}(r) +  G_{n^{\prime}\kappa^{\prime}}(r) \right]
\end{equation}
for the two pseudospin partner states.
These functions are normalized as
\begin{equation}
 \int_{0}^{\infty} dr \: \left| \psi_{2}(n\kappa,r) \right|^{2}
 = \int_{0}^{\infty} dr \: \left| F_{n\kappa}(r) \right|^{2}
\end{equation}
and correspondingly for the second pair. In the case of exact
pseudospin symmetry, the wave functions
$\psi_{2}(n\kappa,r)$ and $\psi_{2}(n^{\prime}\kappa^{\prime},r)$
are identical provided that the upper component wave functions
$F_{n\kappa}(r)$ and $F_{n^{\prime}\kappa^{\prime}}(r)$
in the Dirac spinors are normalized to the same value.
This is beautifully seen in figure \ref{fig:06} 
for the example of section \ref{sec:PS} with a much better agreement
of the two pseudospin partner wave functions than in the
standard approach, cf.\ figure \ref{fig:02}.
In contrast to the wave functions $G_{n\kappa}(r)$ and
$G_{n^{\prime}\kappa^{\prime}}(r)$,  there is no zero in the functions 
$\psi_{2}(n\kappa,r)$ and $\psi_{2}(n^{\prime}\kappa^{\prime},r)$
because they are the lowest states of the supersymmetric partner
Hamiltonians $H_{2}(\kappa)$ and $H_{2}(\kappa^{\prime})$
with the same orbital angular momentum
$l=\tilde{l}=\kappa=-\kappa^{\prime}-1=l^{\prime}$.
The factors $q_{n\kappa}(r)/A_{n\kappa}(r)$ and
$q_{n^{\prime}\kappa^{\prime}}(r)/A_{n^{\prime}\kappa^{\prime}}(r)$ 
in equations (\ref{eq:psi2}) and (\ref{eq:psi2p}) are very
small at small radii $r$ 
explaining the observation of the close similarity of the
wave functions $G_{n\kappa}(r)$ and $G_{n^{\prime}\kappa^{\prime}}(r)$
in the usual comparison. At larger radii, however,
the contribution from the upper component wave functions
$F_{n\kappa}(r)$ and $F_{n^{\prime}\kappa^{\prime}}(r)$, respectively, 
becomes more important.


\subsection{Pseudospin symmetry breaking potential}

The Hamiltonians $H_{2}(\kappa)+E(\kappa)$ and 
$H_{2}(\kappa^{\prime})+E(\kappa^{\prime})$ of the supersymmetric partners
for the pseudospin partner levels are almost identical. The difference
is given by the potential
\begin{eqnarray}
 \label{eq:delta2}
 \Delta_{S}(\tilde{l},r) & = & H_{2}(\kappa)+E(\kappa)
 - \left[ H_{2}(\kappa^{\prime})+E(\kappa^{\prime}) \right]
 \\ \nonumber & = &
 \frac{2}{\sqrt{A}} \frac{d}{dr} 
 \frac{q_{\kappa}-q_{\kappa^{\prime}}}{\sqrt{A}}
 + 2 T \frac{2\kappa+1}{Ar}
\end{eqnarray}
with $\tilde{l} = \kappa = -\kappa^{\prime}-1$ and $A=A_{\kappa}$
where equation (\ref{eq:H2pp}) has been used and the difference between
$A_{\kappa}$ and $A_{\kappa^{\prime}}$ has been neglected.
The quantity $\Delta_{S}(\tilde{l},r)$ 
is a regular function for all radii $r$ 
with the same contribution from the tensor potential $T(r)$
as the previously defined pseudospin-orbit potential $\Delta(\tilde{l},r)$
in (\ref{eq:delta1}). However, the pseudospin breaking potential 
(\ref{eq:delta2}) is a regular function for all $r$ without
a pole as observed for $\Delta(\tilde{l},r)$. 
Without tensor
potential, the pseudospin symmetry breaking potential depends
on the derivative containing the reduced supermomenta of the 
pseudospin partner systems instead on the derivative of the difference $V-S$
as in the traditional approach (\ref{eq:delta1}).
From figure \ref{fig:03} it is evident that $\Delta_{S}(\tilde{l},r)$
with $\tilde{l}=1$ for the example in section \ref{sec:PS}
is really small as compared to $V(r)-S(r)$ and can be considered
as a perturbation. In a relativistic model with tensor potential $T(r)$,
both the non-relativistic spin-orbit potential (\ref{eq:Vso})
and the pseudospin symmetry breaking potential (\ref{eq:delta2})
are directly affected by $T(r)$ with the same functional form,
only the prefactors $\kappa-1$ and $2\kappa+1$ are different.
The dependence of $\Delta_{S}(\tilde{l},r)$ on the scalar and
vector potentials $S(r)$ and $V(r)$ is, however, more subtle through the
Riccati equation for the reduced supermomenta.

\subsection{Potentials in relativistic
Hamiltonians with exact pseudospin symmetry}

In general, there are three potentials, $S(r)$, $V(r)$ and
$T(r)$, in the relativistic Hamiltonian (\ref{eq:H_Dirac}). With one
constraint $\Delta_{S}=0$ to achieve exact pseudospin symmetry,
only two independent functions of these potentials can be specified
freely (subject to some conditions).
Since the difference potential
(\ref{eq:delta2}) is a function of $A(r)$ (containing the sum $V(r)+S(r)$)
and $T(r)$, it is reasonable
to use these quantities as independent variables and to determine the
reduced supermomenta, $q_{\kappa}(r)$ and $q_{\kappa^{\prime}}(r)$ and
finally the scalar and vector potentials, $S(r)$ and $V(r)$.
This approach will be followed in the considerations below.
As an alternative, the potentials in the relativistic Hamiltonian
could be derived from two given supermomenta $q_{\kappa}(r)$ and
$q_{\kappa^{\prime}}(r)$.
In order to simplify the notation, the quantities
\begin{eqnarray}
 \omega & = & \frac{E(\kappa)-E(\kappa^{\prime})}{2\kappa+1}
 \\
  E_{0} & = & \frac{E(\kappa)+E(\kappa^{\prime})}{2}
\end{eqnarray}
are introduced replacing the energy shifts 
$E(\kappa)$ and $E(\kappa^{\prime})$.

From the condition 
$\Delta_{S}(\tilde{l},r) = 0$
one obtains the general solution for the difference
\begin{equation}
 q_{\kappa}(r)-q_{\kappa^{\prime}}(r) = - (2\kappa+1) C(r) 
\end{equation}
with the function
\begin{equation}
 \label{eq:Cdef}
 C(r) = \sqrt{A(r)} \left( D + \int_{0}^{r}
 \frac{T(r^{\prime})}{\sqrt{A(r^{\prime})}} 
 \frac{dr^{\prime}}{r^{\prime}}
 \right)
\end{equation}
where $D$ is an integration constant. Writing
\begin{eqnarray}
 q_{\kappa}(r) & = & q(r) + T(r) - \frac{2\kappa+1}{2} C(r)
 \\
 q_{\kappa^{\prime}}(r) & = & q(r) + T(r) + \frac{2\kappa+1}{2} C(r)
\end{eqnarray}
the difference of the defining equations (\ref{eq:Ric}) for
$q_{\kappa}(r)$ and $q_{\kappa^{\prime}}(r)$ leads to the solution
\begin{equation}
 \label{eq:defq}
    q(r) = \frac{1}{2} \left[
 \frac{(A \omega - C^{\prime}) r-C}{1+Cr} \right] 
 = \frac{1}{2} \left[
 \frac{A \omega r}{1+Cr} - \frac{d}{dr} \ln ( 1+ C r) \right] 
\end{equation}
for the auxiliary momentum $q(r)$. The sum of the two defining equations gives
in combination with the definition of $A(r)$ the 
scalar and vector potentials
\begin{eqnarray}
 \label{eq:resS}
 S(r) & = & \frac{1}{2} \left[ E - E_{0} - A(r) - U(r) \right] + m
 \\
 \label{eq:resV}
 V(r) & = & \frac{1}{2} \left[ E + E_{0} - A(r) + U(r) \right]
\end{eqnarray}
with the auxiliary potential
\begin{eqnarray}
 \label{eq:defW}
 U(r)  & = & V(r)-S(r) + m - E_{0}
 \\ \nonumber 
 & = & \frac{1}{A} \left[ 
 \left( q + T \right) \left( q - T + \frac{1}{r} + \frac{A^{\prime}}{A}\right)
 - q^{\prime} 
 + (2\kappa+1)^{2} \left( \frac{C^{2}}{4} + \frac{C}{2r} \right)
 \right] 
\end{eqnarray}
that is
invariant under the replacement 
$\kappa \leftrightarrow \kappa^{\prime}= -\kappa-1$.
For the determination of the potentials $S(r)$ and $V(r)$ in the 
relativistic Hamiltonian that exhibits exact pseudospin symmetry,
the given functions $A(r)$ and $T(r)$ have to be regular, but
not every choice is admissible.
At least, one has to require that $A(r)>0$ for $r>0$ and that $T(r)$ is 
selected such that the integral in equation
(\ref{eq:Cdef}) is finite for all $r$. In addition,
there is the constraint that $C(r)$ has to be zero at $r=0$ because
$q_{\kappa}(0)=q_{\kappa^{\prime}}(0)=0$ by construction of the 
supersymmetric partner Hamiltonians and that $C(r) \geq 0$, otherwise
the momentum $q(r)$ is not regular for all radii.
It is worthwhile to note that the potential (\ref{eq:defW})
depends on the quantum number
$\kappa$ if $C(r) \neq 0$ and that exact pseudospin symmetry 
is found only in the particular doublet of states with given
$\kappa = \tilde{l} = \kappa$
and $\kappa^{\prime} = -\tilde{l}-1$ if the
relativistic potentials are assumed to be identical for all
states. 

It is instructing to discuss the resulting potentials for
a few particular choices of $A(r)$ and $T(r)$. 
There are two major cases depending on the value of $A(r)$ at $r=0$.

\begin{enumerate}

\item $A(0) = 0$ but $A(r) >0$ if $r>0$. In this case, which might not be 
realized in a physical system, the integration 
constant $D$ in equation (\ref{eq:Cdef}) 
can have any non-negative value as long
as $C(r) \geq 0$ for all $r$. If $C(r)$ is not identically zero,
the reduced
supermomenta of the pseudospin partner system with
different $\tilde{l}$ are different
and the potential $U(r)$ explicitly depends on $\kappa$.
This can occur even for vanishing tensor potential if $D>0$.


\item $A(0) \neq 0$, the case usually encountered in
physical systems. It follows that $D=0$ and that $C(r) \neq 0$ 
only if there is a tensor interaction. 
See the discussion in the paragraph above. 

\end{enumerate}

In most relativistic Hamiltonians 
without a tensor interaction and everywhere finite effective
mass of the nucleon there is still a large variation of potentials
that exhibit exact pseudospin symmetry.
For $T(r)=0$ and then $C(r) = 0$
for all radii $r$, the reduced supermomenta of the pseudospin partners
are identical with
\begin{equation}
 q_{\kappa}(r) = q_{\kappa^{\prime}}(r) = q(r)
 = \frac{A(r)}{2} \omega r 
\end{equation}
and the auxiliary potential becomes
\begin{equation}
 U(r) = \frac{1}{A} \left[ q^{2} +\frac{q}{r} 
 +  \frac{A^{\prime}(r)}{A(r)} q - q^{\prime}\right]
 =  \frac{A(r)}{4} \omega^{2} r^{2} \: ,
\end{equation}
i.e., a modified harmonic oscillator potential with position depending
effective mass.  With the scalar and vector potentials
(\ref{eq:resS}) and (\ref{eq:resV}), respectively, the 
central and and spin-orbit potentials (\ref{eq:Vc}) and (\ref{eq:Vso})
are found as
\begin{eqnarray}
 V_{c}(r) & = & E_{0}-m + U(r)
 + \frac{1}{2} V_{so}(r)
 \\
 V_{so}(r) & = & - \frac{2A^{\prime}}{A^{2}r} 
\end{eqnarray}
in the non-relativistic Hamiltonian (\ref{eq:HFnr}).
The standard harmonic oscillator is obtained
for constant $A(r)=2m$, $T(r)=0$ and $E(\kappa) = m+(2\kappa+1)\omega/2$
such that $E_{0} = m$. Then the scalar and vector potentials are
\begin{equation}
 S(r) = - V(r) = - \frac{m}{4} \omega^{2} r^{2}  
\end{equation}
leading to the non-relativistic central and spin-orbit potentials
\begin{equation}
 V_{c}(r) = \frac{m}{2} \omega^{2} r^{2}\qquad
 V_{so}(r) = 0 \: .
\end{equation}
In general, the choice of the position depending
effective mass $m_{\rm eff}(r)= A_{\kappa}(r)/2$
has an important effect on the potentials
that appear in the non-relativistic Hamiltonian. 
However, a reasonable effective mass
should approach the free nucleon mass $m$ for large radii $r$. Then
the potentials will approach the harmonic oscillator form in the limit
$r \to \infty$.

\section{Conclusions and outlook}

The occurrence of almost degenerate pseudospin partner levels
in atomic nuclei is usually attributed to a
symmetry of the underlying relativistic Hamiltonian as it appears, 
e.g., in RMF models. The lower-component radial wave functions in the
Dirac spinors are found to be very similar. From the differential equation
of these wave functions a pseudospin-orbit potential is
extracted that breaks the pseudospin symmetry. However, it is not a
regular function and it cannot be considered as a small perturbation.
In addition, this potential does not vanish for the usual harmonic
oscillator potential which exhibits exact pseudospin symmetry.

In the supersymmetric approach to pseudospin symmetry, the supersymmetric
partner Hamiltonians of the upper-component wave functions
are compared. It leads to a regular symmetry breaking potential that
is small in comparison with the dominating central potential in the mean-field
Hamiltonian. The corresponding partner wave functions are identical
in the case of exact pseudospin symmetry. 
Employing both exact and broken supersymmetry, all levels
in a sequence of states
for given pseudo orbital angular momentum $\tilde{l}>0$, including the
lowest unpaired state, fit into the scheme. This lowest level
defines the reduced
supermomentum and consequently the potentials in the Hamiltonians
for both pseudospin partner systems in the case of exact pseudospin
symmetry. All higher lying pairs are degenerate in energy.
From the condition of vanishing symmetry breaking potential, general
relativistic potentials that exhibit exact pseudospin symmetry can be derived.
Modified harmonic oscillator potentials with position depending
effective mass are the most simple class.

The present description relies on the differential equations
for the radial wave functions and the corresponding first-order
differential operators in the factorization of the Hamiltonian for
given quantum number $\kappa$. It is also possible to generalize the
approach to a form independent of a particular $\kappa$.
In this case the operators 
\begin{equation}
 B^{+} = \left[ q - \vec{\sigma} \cdot \left( \vec{T} - i \vec{p} 
\right) \right] \frac{1}{\sqrt{A}}
 \qquad
 B^{-} = \frac{1}{\sqrt{A}} \left[ q - \vec{\sigma} \cdot 
\left( \vec{T} + i \vec{p} \right) \right] 
\end{equation}
in the generators (\ref{eq:supercharge}) of the symmetry transformation
act on the full wave functions in coordinate space.
These and more formal aspects of the supersymmetric approach can be studied
in the future.

The symmetry breaking potential
depends on derivatives of the supermomenta for the two pseudospin
partner states and the tensor potential. 
The supermomenta itself are determined via a Riccati
equation by the scalar, vector and tensor potentials in the relativistic
Hamiltonian. A simple example shows that the reduced supermomenta of
pseudospin partner states are very similar rising approximately
linear at small radii as in the case of the harmonic oscillator.
It seems that all saturating potentials with a flat bottom will
show an approximate pseudospin symmetry. This question can be explored
in more detail by expanding the reduced supermomenta in a power series
in terms of the radius $r$.

Calculating the supermomenta for actual selfconsistent RMF models will help to
quantify the various sizes of the observed pseudospin symmetry breaking.
The study of the symmetry breaking potential will eventually lead
to improved RMF parametrizations since it can help to identify
the relevant modifications of the model that are required to obtain
a better description of the experimentally observed breaking and 
restoration of the pseudospin symmetry in nuclei.
It will be interesting to see how the isospin dependence of the effective
interaction changes the degree of pseudospin symmetry breaking.
Of special importance is the tensor interaction (neglected in most
RMF models) that affects both the
spin-orbit and the pseudospin-orbit splitting.

The method that is developed in this paper is not
restricted to a relativistic description of the single-particle
states. In principle, it can also be applied to the usual
Schr\"{o}dinger equation for the single-particle wave functions
as it appears, e.g., in the Skyrme Hartree-Fock method, because
it relies only on the Schr\"{o}dinger-equivalent differential
equation of the upper component in the Dirac spinor.
Instead of the scalar, vector and tensor potentials in
the relativistic Hamiltonian, there are also at least three independent
quantities that characterize the mean-field in the non-relativistic
Hamiltonian: the position depending effective mass, 
the central and the spin-orbit potential. The relation between these
functions and the reduced supermomentum will be different as compared
to the relativistic description, however, the same principles
of the supersymmetric method can be applied.

\section*{Acknowledgements}
The author is grateful to O. Sorlin and D. Lacroix for many 
stimulating discussions and a critical reading of the manuscript.

\section*{Note added in proof}
Pseudospin symmetry was found as a particular limit in the application
of supersymmetry to the Dirac equation in Refs. [Lev04,Lev05] by directly
factorizing the Dirac Hamiltonian.

\newpage

\section*{Tables}

\begin{table}[h]
\caption{\label{tab:kjl}%
Relation between the various quantum numbers in the relativistic 
description of single-particle states. Pseudospin partners with the same
pseudoangular momentum $\tilde{l}$ are placed in the same column.}
\begin{tabular}{cccccccccc}
  \hline \hline
  $\tilde{l}$ & 0 & 1 & 2 & 3 & 4 & 5 & 6 & 7 & \dots \\
  \hline \hline
  $\kappa$    &   & 1 & 2 & 3 & 4 & 5 & 6 & 7 & \dots \\
  \hline
  $j^{\pi}$   &   & $1/2^{+}$ & $3/2^{-}$ & $5/2^{+}$ 
  & $7/2^{-}$ & $9/2^{+}$ & $11/2^{-}$ & $13/2^{+}$ & \dots \\
  $l$         &   & 0 & 1 & 2 & 3 & 4 & 5 & 6 & \dots \\
  \hline \hline
  $\kappa$    & $-1$ & $-2$ & $-3$ & $-4$ & $-5$ & -6 & -7 & -8 & \dots \\
  \hline
  $j^{\pi}$   & $1/2^{-}$ & $3/2^{+}$ & $5/2^{-}$ & $7/2^{+}$ & $9/2^{-}$ &
  $11/2^{+}$ & $13/2^{-}$ & $15/2^{+}$ & \dots \\
  $l$         & 1 & 2 & 3 & 4 & 5 & 6 & 7 & 8 & \dots \\
  \hline \hline
\end{tabular}
\end{table}

\newpage

\section*{Figures}

\end{document}